# AI Tool Adoption and Loyalty among Product, Industrial, and UX/UI Designers across the Double Diamond

**Sepideh Tajarmakan** *

[1]Independent Researcher

m.s.tajarmakan@gmail.com
ORCID 0009-0009-3418-8893

**Khashayar Hojjati Emami**[2]

[2]Iran University of Art, Department of Industrial Design

k.emami@art.ac.ir
ORCID 0000-0002-7325-8091

** Corresponding author*

**ABSTRACT**

Designers share many AI tools, yet their roles produce different work. This survey compared 337 Product, Industrial and UX/UI Designers from 40 countries and regions across the four phases of the Double Diamond. It measured AI adoption, breadth and frequency, tool function and the number of tools named per phase. A tool was loyal when a role picked it in all four phases. Industrial Designers adopted AI less often (73.9% against 88.2% and 87.9%), a gap that went with physical projects once project type and country were held constant. Of 60 AI tools, 25 were common to all three roles and carried at least 90% of each role's picks. Tool function differed by role, most of all in Develop. A few loyal tools carried about half of the picks, led by ChatGPT. Tool counts fell across the phases in every role. A hurdle model found a Phase × Role interaction for using AI in a phase ($p = 0.041$), which was not shown for the number of tools ($p = 0.080$). Predicted adoption among UX/UI Designers fell from 87.2% in Define to 43.9% in Develop. Support for designers needs one shared core, role-specific tools and most attention in Develop.

# INTRODUCTION

AI has become part of most designers' projects. In a survey of 443 designers, 79.7% reported AI use in one or more phases of the Double Diamond. General conversational tools led three of its four phases (Tajarmakan & Hojjati Emami, 2026). The Double Diamond gives designers one four-phase map from Discover to Deliver (Design Council, 2007), and it is widely used to place AI in design work (Y. Zhou & Chen, 2025; Mirzaei et al., 2026). Yet the designers on this map do not all hold the same job.

The names of design roles overlap. Industrial design is defined as work on products, systems, services and experiences (World Design Organization, 2015). UX work deals with user experience, a person's perceptions and responses from using a system (International Organization for Standardization [ISO], 2019). The title product designer first named designers of physical products and was later taken up by designers of websites and apps (Chochinov, 2023). Among 692 UX professionals, it was the second most common title (Rosala & Krause, 2019). A role name alone may thus say little about how a designer uses AI.

Most evidence comes from small groups or from stated intentions. Interviews and workshops covered between 10 and 32 designers or students each, mostly within a single role (J. Li et al., 2024; Uusitalo et al., 2024; Abrahamsen & Sjödell, 2026; Süner-Pla-Cerdà et al., 2026). Larger surveys modelled intention to adopt AI or sorted designers into adopter types (Zou et al., 2026; Fang, Guo, et al., 2026). One compared 607 design professionals in China and the United Kingdom without separating disciplines, and called for studies by design area (Fang, Zhang, et al., 2026). Our earlier analysis of the same survey reported tools and intensity by phase for all roles together (Tajarmakan & Hojjati Emami, 2026). Research on continued use follows one system over time (Bhattacherjee, 2001), not the set of tools a designer keeps from phase to phase. A direct comparison of roles on reported use is still missing.

This study compares Product, Industrial and UX/UI Designers on AI adoption, tool choice and intensity across the Double Diamond. It draws on 337 designers from 40 countries and regions (Tajarmakan & Hojjati Emami, 2026). The results can show tool makers, educators and design teams where one AI toolkit serves all three roles and where it does not. Three research questions follow.

**RQ1 — Adoption:** Do AI adoption, breadth and frequency differ across Product, Industrial and UX/UI Designers?

**RQ2 — Tools:** Which AI tools, by output and function, does each role use, and which tools does each role keep in all four phases?

**RQ3 — Intensity:** How does the number of AI tools named change across the four phases, and does this path differ by role?

# LITERATURE REVIEW

## *Design Roles and AI*

Research on AI in design has mostly looked at one role at a time. Among 20 UX designers, experienced practitioners saw generative AI as an assistant, while juniors risked losing skills (J. Li et al., 2024). Twelve UX practitioners who used ChatGPT for four to six weeks found it useful for guidelines, user profiles and simulated stakeholders, yet weak at prototyping (Z. Zhou et al., 2024). In workshops with 32 industrial designers and students, image generators set automation against designerly autonomy (Süner-Pla-Cerdà et al., 2026). Twenty expert industrial designers reported pressure for speed and a risk of handing small decisions to the tool (Abrahamsen & Sjödell, 2026). Among 327 product designers, intention to use AI came from combinations of factors such as usefulness, task fit and organizational support (H. Li et al., 2025). Interviews across UX and industrial design tied attitudes to agency and control (Uusitalo et al., 2024). These studies compared views within roles, and the phase of work offers a second lens on where roles may part.

### *AI Tools across the Design Phases*

AI tools differ in what they produce and in where they help. A review grouped generative models by the medium they produce, with text, image and 3D models among the groups (Gozalo-Brizuela & Garrido-Merchan, 2023). The Double Diamond alternates divergent and convergent phases, once to frame the problem and once to shape its answer (Design Council, 2007), although real workflows bend to the situation (Suchman, 1987; Retelny et al., 2017). Systems built to support design cluster in Develop and Deliver (Lee et al., 2025), and a qualitative review found a different role for AI at each stage (Mirzaei et al., 2026). Among 20 novice designers, AI helped most with ideas and problem definition, while evaluation stayed human-led (Chen et al., 2025). Reported use showed conversational tools ahead in Discover, Define and Deliver and AI-native image generation ahead in Develop (Tajarmakan & Hojjati Emami, 2026). This evidence pooled all roles and counted tools phase by phase, so the tools a designer keeps across phases need their own measure.

### *Measuring Use and Loyalty*

Use can be measured in more than one way. Usage links a user, a system and a task, and its measure should fit the question asked (Burton-Jones & Straub, 2006). Acceptance research predicts intention from perceived usefulness (Davis, 1989), and a later model added habit (Venkatesh et al., 2012). Continued use rests on satisfaction and usefulness (Bhattacherjee, 2001), and habit grows from frequent past use until use becomes automatic (Limayem et al., 2007). Design studies have measured frequency for one tool category (Tang et al., 2024) or treated AI use as one general behavior (Y. Lu et al., 2024). Marketing research calls the repeated choice of one brand or set of brands loyalty, but warns that repeat patterns alone do not show commitment (Oliver, 1999). People also combine tools with overlapping functions, and the role of each tool shifts over time (Bødker & Klokmose, 2012). To our knowledge, no study has traced which AI tools designers keep across all phases of one project, a question that needs tool names for every phase.

## METHOD

### *Study Design and Participants*

This cross-sectional survey followed AI use through the four phases of the Double Diamond (Design Council, 2007). Data came from one online questionnaire in English (Google Forms), open from August 2025 to February 2026. The same survey supplied the data of companion paper (Tajarmakan & Hojjati Emami, 2026). Every question referred to the designer's most recent project. Designers were reached through purposive, convenience and snowball sampling in professional networks and design groups on LinkedIn, Facebook and Telegram.

Of the 443 respondents, this paper compares the three largest groups by primary role (Q6). Product Designers (n = 119), Industrial Designers (n = 119) and UX/UI Designers (n = 99) formed a sample of 337 designers from 40 countries and regions. Graphic designers (n = 40), other roles (n = 35) and design students (n = 31) formed smaller groups and were left out. Recruitment set no quota by role.

## *Questionnaire and Measures*

The form had 70 items in nine sections (Table 1). Figure 1 shows the route through the form. After consent and eight background items (Q2 to Q9), one question asked whether the designer used AI in design work (Q10). Of the 337 designers, 284 answered "Yes" and 53 answered "No". Those who used AI reported how often (Q11) and in which steps of their work (Q12).

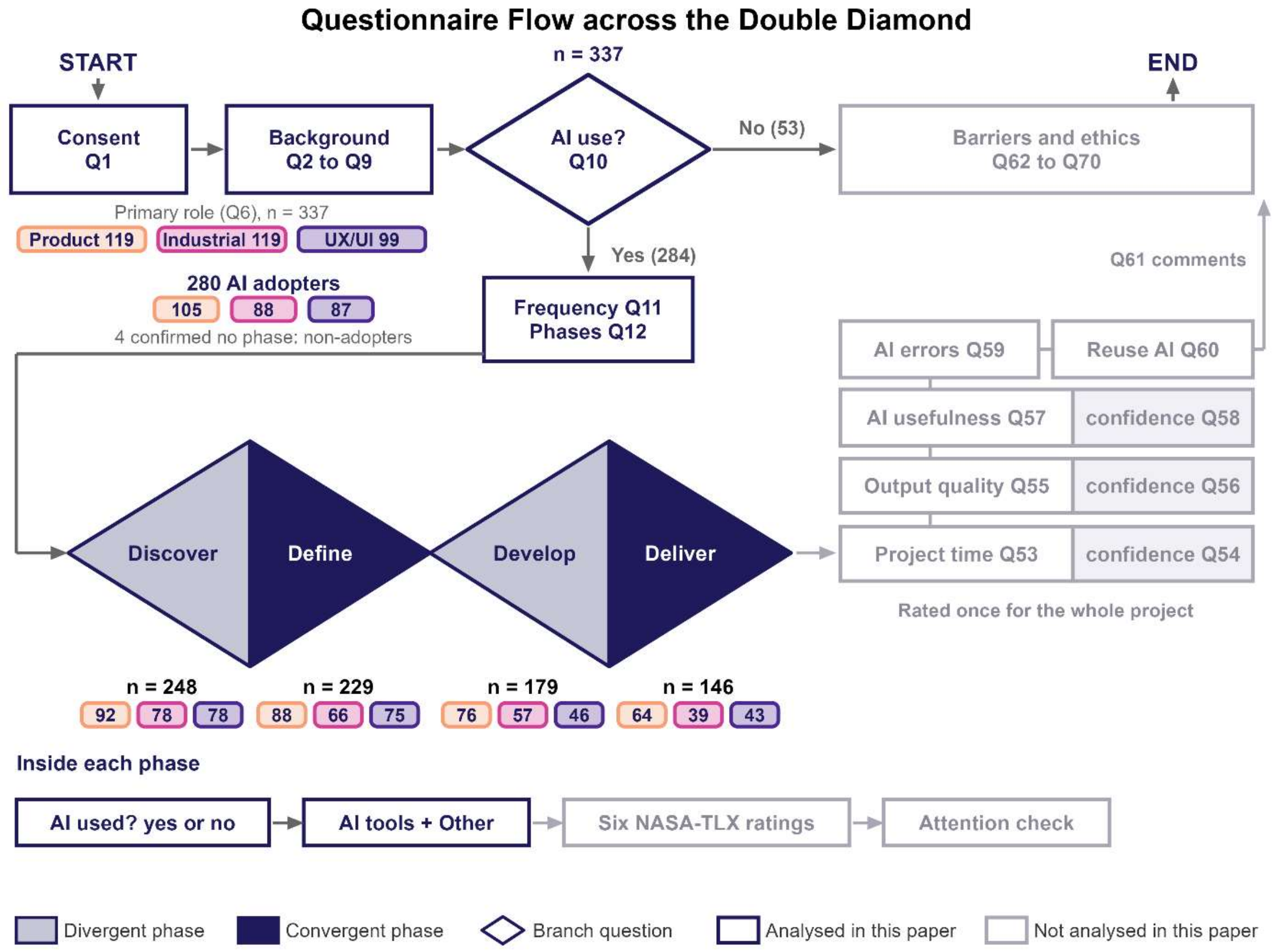

***Note. Route through the 70-item questionnaire for the 337 designers of this paper. Role chips give n for Product, Industrial and UX/UI Designers. Grey boxes were not analyzed in this paper.***

**Figure 1: Questionnaire Flow across the Double Diamond**

Four phase blocks followed in the order of the Double Diamond. Each began by asking whether AI was used in that phase (Q13, Q23, Q33, Q43). After a "Yes", the designer ticked AI tools from a list or typed their names under "Other" (Q14 to Q15, Q24 to Q25, Q34 to Q35, Q44 to Q45).

The lists held seven, six, six and five tools (Table 2). Workload ratings, the KPI block and the barriers block serve the companion papers and are not analyzed here.

**Table 1: Questionnaire Sections and Item Counts**

| Section | Question range | Items |
|---|---|---|
| Consent | Q1 | 1 |
| Demographics | Q2 – Q9 | 8 |
| AI overall | Q10 – Q12 | 3 |
| Discover phase (AI-tools, NASA-TLX) | Q13 – Q22 | 10 |
| Define phase (AI-tools, NASA-TLX) | Q23 – Q32 | 10 |
| Develop phase (AI-tools, NASA-TLX) | Q33 – Q42 | 10 |
| Deliver phase (AI-tools, NASA-TLX) | Q43 – Q52 | 10 |
| KPI | Q53 – Q61 | 9 |
| Ethics and Barriers | Q62 – Q70 | 9 |
| Total | | **70** |

***Note. Phase rows are indented under the phase-by-phase section. Each phase block was shown only to participants who answered "Yes" to the general screening question (Q10).***

**Tick lists shaped the counts.** Offered tools carried most picks in every phase, from 69.7% in Develop to 85.4% in Deliver (Table 2). ChatGPT was offered in Discover, Define and Deliver. All 22 of its Develop picks were typed in under "Other" (Product 10, Industrial 5, UX/UI 7). Phase differences in tool use may thus partly follow the form.

**Table 2: AI Tools Offered in Each Phase and Picks of Offered Tools**

| Phase | Tools offered in the form | Picks of offered tools n (%) |
|---|---|---|
| **Discover** | ChatGPT, Perplexity AI, Elicit, Figma AI, Miro AI, Autodesk Generative Design, Adobe Sensei | 402 (77.3) |
| **Define** | ChatGPT, Notion AI, FigJam AI, Miro AI, Autodesk Fusion 360 AI, Whimsical AI | 354 (84.5) |
| **Develop** | Midjourney, DALL·E, Figma AI, Adobe Firefly, Runway ML, Autodesk Fusion 360 AI | 253 (69.7) |
| **Deliver** | Adobe Photoshop, Canva AI, Runway ML, ChatGPT, Descript | 228 (85.4) |

***Note. Offered = a tick box in that phase (Q14, Q24, Q34, Q44). ChatGPT was not offered in Develop***

**Adoption rested on the phase questions.** A phase AI adopter answered "Yes" for that phase, and an AI adopter did so in at least one phase. Four designers answered "Yes" to Q10 and ticked a step in Q12, yet answered "No" to every phase question. They counted as non-adopters, which left 280 AI adopters. Phase AI adopters numbered 248 in Discover, 229 in Define, 179 in Develop and 146 in Deliver. The phase questions found more AI use than the general step item in all four phases (McNemar $\chi^2(1) = 4.21$ to 39.56, Appendix Table 1; McNemar, 1947).

**Three measures described AI use**. **Breadth** was the number of phases with AI use, from 1 to 4. **Frequency** was the general frequency of AI use (Q11), from 1 (once a month or less) to 5 (daily). **Intensity** was the number of AI tools a designer named in a phase, counted for all 337 designers with 0 for no use. One tool pick meant one designer naming one tool in one phase.

## *Data Cleaning and Tool Classification*

Tool names were cleaned in Excel and matched to the AI tool taxonomy of Paper 1 on letters and digits only (Tajarmakan & Hojjati Emami, 2026). The taxonomy held 93 entries, of which 73 were AI tools. Entries without an AI feature or without one identifiable tool, such as internal company software, did not count as tools. The three roles named 60 AI tools in 1,569 picks (Product 620, Industrial 441, UX/UI 508). Of these tools, 18 (30.0%) appeared in at least one tick list and 42 (70.0%) were typed in. Eleven tools (18.3%) were picked in all four phases, and 26 (43.3%) in only one. Appendix Table 2 shows the coding. Each tool was classified by its main output and by its function within that output (Gozalo-Brizuela & Garrido-Merchan, 2023; Q. Lu et al., 2024; Shi et al., 2024; Doshi & Moore, 2026). Five outputs held ten functions (Figure 3). Two coders agreed on the output ($\kappa = 1.00$) and the function ($\kappa = 0.937$) of each tool, a near-perfect level (Landis & Koch, 1977; Tajarmakan & Hojjati Emami, 2026). Some phase AI adopters named no identifiable AI tool. They ticked "Other" without a name, named an internal company tool or named a tool without AI features. This held for two designers in Discover, one in Define, 12 in Develop and two in Deliver (17 of 802 phase answers). They stayed in the analyses with an intensity of 0.

## Data Analysis Plan

Analyses followed the three research questions. Intensity counts were skewed (Shapiro-Wilk $W = 0.72$ to $0.87$ per phase, all $p < 0.001$; Shapiro & Wilk, 1965), so we used rank tests and count models. For **RQ1**, a chi-square test compared adoption across roles, with Cramér's V as effect size (Pearson, 1900; Cohen, 2013). Kruskal-Wallis tests compared breadth and frequency (Kruskal & Wallis, 1952). Spearman correlations linked frequency with breadth within each role (Spearman, 1904). A logistic regression then added project type, experience, education and country (Iran or other) to role (Appendix Table 4). For **RQ2**, chi-square tests compared tool functions across roles, and AI-native with embedded tools. Fisher's exact tests with 10,000 Monte Carlo draws checked tables with small expected counts (Fisher, 1922). One designer gave several picks, so a designer-level permutation test kept each designer's picks together. It shuffled role labels across designers 10,000 times. A tool was common in a phase when all three roles picked it there, shared when two roles did and exclusive when one did. Loyal tools were those a role picked in all four phases. For **RQ3**, Friedman tests compared intensity across the four phases within each role, with Kendall's W as effect size (Friedman, 1937; Kendall & Smith, 1939). Wilcoxon signed-rank tests then compared each pair of phases (Wilcoxon, 1945). Kruskal-Wallis tests compared the roles within each phase. A mixed-effects hurdle model, fit with glmmTMB (Brooks et al., 2017), split adoption from the number of tools. Part 1 modelled AI use in all 1,348 designer-phase rows, and Part 2 the number of tools in the 785 rows with at least one named tool. Both parts held a Phase × Role interaction and a random intercept per designer, tested by a likelihood ratio test. Holm correction controlled repeated tests (Holm, 1979). The phase tool files were merged in Python 3.13 with pandas (Python Software Foundation, 2024). Analyses ran in R 4.5.3 (R Core Team, 2026; packages openxlsx, ggplot2, gridExtra, patchwork and glmmTMB).

## Validity and Reliability

Validity and reliability followed Runeson and Höst (2009), as in the companion paper (Tajarmakan & Hojjati Emami, 2026). **Construct validity.** The instrument and tool taxonomy were built from a literature and market review of AI tools in design practice. Three expert professors rated every item of the full questionnaire as relevant before data collection (item- and scale-level content validity index 1.00; Davis, 1992; Polit & Beck, 2006). Loyalty was not asked directly. It was inferred from the tools designers named in each phase, so it is a behavioral proxy for commitment and not a measure of attitude (Oliver, 1999). Every question referred to the most recent project, which tied answers to one instance and reduced recall error. **Internal validity.** The design is cross-sectional and correlational. Role differences are associations and not causal effects. Project type, experience, education and country entered the adjusted models as controls. **External validity.** Snowball sampling gives a non-probability sample. Iran supplied 145 of the 337 designers (43.0%), and the survey was in English only. Results therefore describe this sample and may not extend to all designers. **Reliability.** Google Forms allowed one response per account and required every shown item, which prevented duplicate and incomplete responses. Two coders classified each tool by output and function and agreed almost perfectly (Data Cleaning and Tool Classification). Tool names were matched to the taxonomy by a fixed rule, and all steps were scripted, so the analyses can be repeated from the deposited files.

# RESULTS

## *Demographics*

Table 3 describes the three roles. Most designers were aged 20 to 29 (198 of 337, 58.8%), and UX/UI Designers were the youngest group (65.7%). UX/UI Designers also had the least experience, since almost half had worked for one to three years (49.5%). Product Designers held the largest share with seven or more years of experience (39.5%).

**Table 3: Demographic Information by Role (N = 337)**

| | | **Product (n = 119)** | | **Industrial (n = 119)** | | **UX/UI (n = 99)** | | **Total (N = 337)** | |
|---|---|---|---|---|---|---|---|---|---|
| **Variable** | **Category** | **n** | **%** | **n** | **%** | **n** | **%** | **n** | **%** |
| **Age** | Under 20 | 1 | 0.8 | 1 | 0.8 | 1 | 1.0 | 3 | 0.9 |
| | 20 to 29 | 60 | 50.4 | 73 | 61.3 | 65 | 65.7 | 198 | 58.8 |
| | 30 to 39 | 50 | 42.0 | 33 | 27.7 | 28 | 28.3 | 111 | 32.9 |
| | 40 to 49 | 5 | 4.2 | 10 | 8.4 | 5 | 5.1 | 20 | 5.9 |
| | 50 to 59 | 3 | 2.5 | 1 | 0.8 | 0 | 0.0 | 4 | 1.2 |
| | 60 and above | 0 | 0.0 | 1 | 0.8 | 0 | 0.0 | 1 | 0.3 |
| **Gender** | Female | 33 | 27.7 | 36 | 30.3 | 47 | 47.5 | 116 | 34.4 |
| | Male | 62 | 52.1 | 64 | 53.8 | 38 | 38.4 | 164 | 48.7 |
| | Non-binary | 0 | 0.0 | 2 | 1.7 | 1 | 1.0 | 3 | 0.9 |
| | Prefer not to say | 24 | 20.2 | 17 | 14.3 | 13 | 13.1 | 54 | 16.0 |
| **Education** | High school | 0 | 0.0 | 5 | 4.2 | 3 | 3.0 | 8 | 2.4 |
| | Bachelor | 46 | 38.7 | 62 | 52.1 | 54 | 54.5 | 162 | 48.1 |
| | Master | 60 | 50.4 | 45 | 37.8 | 39 | 39.4 | 144 | 42.7 |
| | Doctorate | 10 | 8.4 | 7 | 5.9 | 3 | 3.0 | 20 | 5.9 |
| | Other | 3 | 2.5 | 0 | 0.0 | 0 | 0.0 | 3 | 0.9 |
| **Experience** | Less than 1 year | 6 | 5.0 | 10 | 8.4 | 8 | 8.1 | 24 | 7.1 |
| | 1 to 3 years | 27 | 22.7 | 43 | 36.1 | 49 | 49.5 | 119 | 35.3 |
| | 4 to 6 years | 39 | 32.8 | 27 | 22.7 | 29 | 29.3 | 95 | 28.2 |
| | 7+ years | 47 | 39.5 | 39 | 32.8 | 13 | 13.1 | 99 | 29.4 |
| **Project type** | Digital | 70 | 58.8 | 15 | 12.6 | 83 | 83.8 | 168 | 49.9 |
| | Hybrid | 30 | 25.2 | 47 | 39.5 | 15 | 15.2 | 92 | 27.3 |
| | Physical | 19 | 16.0 | 57 | 47.9 | 1 | 1.0 | 77 | 22.8 |
| **Most recent project** | Today | 57 | 47.9 | 51 | 42.9 | 49 | 49.5 | 157 | 46.6 |
| | In the past week | 45 | 37.8 | 29 | 24.4 | 24 | 24.2 | 98 | 29.1 |
| | In the past month | 7 | 5.9 | 21 | 17.6 | 14 | 14.1 | 42 | 12.5 |
| | In the past 3 months | 7 | 5.9 | 10 | 8.4 | 7 | 7.1 | 24 | 7.1 |
| | More than 3 months | 3 | 2.5 | 8 | 6.7 | 5 | 5.1 | 16 | 4.7 |

***Note. n = designers. % = share of the role. The gender item was added during data collection, so "Prefer not to say" also holds designers who were never asked.***

Project type split the roles most clearly. Industrial Designers worked mostly on physical (47.9%) or hybrid projects (39.5%), while UX/UI Designers worked almost only on digital ones (83.8%). Product Designers had a mixed profile (58.8% digital). The gender item was added during data collection, so the "Prefer not to say" group (13.1% to 20.2%) also holds designers who were never asked Product Designers came from 22 countries and regions, Industrial Designers from 26 and UX/UI Designers from 23 (Appendix Table 3). After Iran, India and the United States led among Product Designers (nine each). India, Italy and Turkiye followed among Industrial Designers. UX/UI Designers had the most even spread of the three roles, with Italy, India and Turkiye after Iran. Project type and country both entered the adjusted model of adoption.

## *AI Adoption and Breadth by Role*

The first research question compared AI adoption, breadth and frequency across the three roles. **Adoption** differed by role ($\chi^2(2, N = 337) = 10.93$, $p = 0.004$, $V = 0.18$). Product Designers (105 of 119, 88.2%) and UX/UI Designers (87 of 99, 87.9%) adopted AI more often than Industrial Designers (88 of 119, 73.9%). Role overlapped with project type, since only one UX/UI Designer worked on a physical product (Table 3). In the adjusted model ($n = 334$), role was no longer significant once project type and country were held constant ($\chi^2(2) = 3.37$, $p = 0.185$, Appendix Table 4). Project type remained significant ($\chi^2(2) = 9.38$, $p = 0.009$), with lower odds for physical than for digital projects (OR = 0.25, 95% CI [0.10, 0.63]). Designers in Iran adopted AI less often than designers elsewhere (OR = 0.45, 95% CI [0.24, 0.83], $p = 0.010$). Experience and education showed no significant link ($p = 0.119$ and $p = 0.112$). The lower adoption of Industrial Designers thus went with their projects, though the overlap limits how far the two can be told apart.

**Table 4: Workflow Breadth and Phase Combinations among Confirmed Adopters, by Role (n = 280)**

| | Product (n = 105) | | Industrial (n = 88) | | UX/UI (n = 87) | | Total (n = 280) | |
|---|---|---|---|---|---|---|---|---|
| **Breadth and phase combination** | **n** | **%** | **n** | **%** | **n** | **%** | **n** | **%** |
| **1 phase** | **9** | **8.6** | **11** | **12.5** | **9** | **10.3** | **29** | **10.4** |
| Discover | 4 | 44.4 | 3 | 27.3 | 4 | 44.4 | 11 | 37.9 |
| Define | 2 | 22.2 | 4 | 36.4 | 3 | 33.3 | 9 | 31.0 |
| Develop | 3 | 33.3 | 2 | 18.2 | 2 | 22.2 | 7 | 24.1 |
| Deliver | 0 | 0.0 | 2 | 18.2 | 0 | 0.0 | 2 | 6.9 |
| **2 phases** | **22** | **21.0** | **26** | **29.5** | **29** | **33.3** | **77** | **27.5** |
| Discover + Define | 14 | 63.6 | 14 | 53.8 | 19 | 65.5 | 47 | 61.0 |
| Discover + Develop | 3 | 13.6 | 10 | 38.5 | 3 | 10.3 | 16 | 20.8 |
| Develop + Deliver | 4 | 18.2 | 1 | 3.8 | 0 | 0.0 | 5 | 6.5 |
| Define + Develop | 1 | 4.5 | 0 | 0.0 | 3 | 10.3 | 4 | 5.2 |
| Discover + Deliver | 0 | 0.0 | 0 | 0.0 | 3 | 10.3 | 3 | 3.9 |
| Define + Deliver | 0 | 0.0 | 1 | 3.8 | 1 | 3.4 | 2 | 2.6 |
| **3 phases** | **29** | **27.6** | **27** | **30.7** | **21** | **24.1** | **77** | **27.5** |
| Discover + Define + Develop | 14 | 48.3 | 16 | 59.3 | 10 | 47.6 | 40 | 51.9 |
| Discover + Define + Deliver | 9 | 31.0 | 7 | 25.9 | 11 | 52.4 | 27 | 35.1 |
| Discover + Develop + Deliver | 3 | 10.3 | 4 | 14.8 | 0 | 0.0 | 7 | 9.1 |
| Define + Develop + Deliver | 3 | 10.3 | 0 | 0.0 | 0 | 0.0 | 3 | 3.9 |
| **4 phases** | **45** | **42.9** | **24** | **27.3** | **28** | **32.2** | **97** | **34.6** |
| Discover + Define + Develop + Deliver | 45 | 100.0 | 24 | 100.0 | 28 | 100.0 | 97 | 100.0 |
| **All AI adopters** | **105** | **100.0** | **88** | **100.0** | **87** | **100.0** | **280** | **100.0** |

***Note. Bold rows give the number of phases with AI use, as % of the role's AI adopters (105, 88 and 87). Indented rows give the phases used together, as % of the role's AI adopters at that breadth.***

Among the 280 AI adopters, **breadth** differed by role ($H(2) = 6.02$, $p = 0.049$, Table 4). Product Designers used AI in the most phases ($M = 3.05$), against 2.73 for Industrial and 2.78 for UX/UI

Designers. They also had the largest share in all four phases (45 of 105, 42.9%). Discover + Define was the most common pair at breadth 2 in every role. At breadth 3, Product and Industrial Designers most often added Develop to this pair, while UX/UI Designers split almost evenly between Deliver (11 of 21) and Develop (10 of 21). The one-phase groups held only nine to 11 designers per role, so Table 4 describes them without comparison.

**Frequency** of AI use also differed by role ($H(2) = 18.93$, $p < 0.001$). Daily use was reported by 53 of 87 UX/UI Designers (60.9%) and 57 of 105 Product Designers (54.3%), against 25 of 88 Industrial Designers (28.4%). In every role, frequency rose with breadth ($\rho$ = 0.315, 0.241 and 0.370, p = 0.001, 0.024 and <0.001). Figure 2 crosses frequency with breadth. Daily use was rare among one-phase Product Designers (1 of 9) and most common at breadth 4 (31 of 45). Industrial Designers peaked at “A few times a week” (17 at breadth 3). Among UX/UI Designers, daily use was common at every breadth from 2 to 4 (15, 16 and 21 designers). Role was thus linked to adoption, breadth and frequency, though the adoption gap went with project type in the adjusted model.

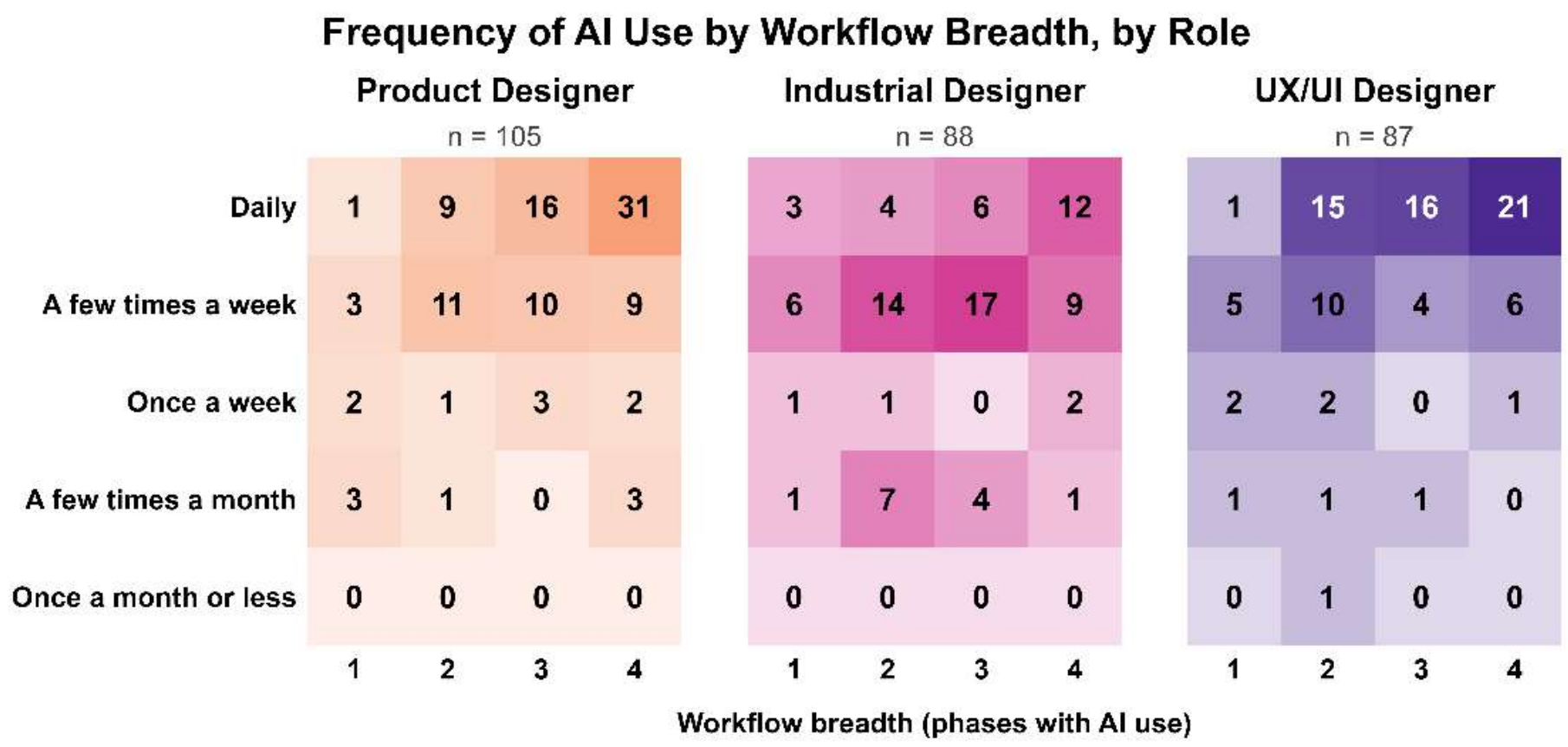

***Note. Cells count AI adopters by frequency (Q11) and breadth (n = 105, 88 and 87). Each role has its own colour and scale, with the largest cell in full colour, so colour strength is relative within each role.***

**Figure 2: Frequency of AI Use by Workflow Breadth, by Role**

## *Tool Function by Role*

The second research question compared the AI tools of the three roles. AI systems have been grouped by their level of automation (Parasuraman et al., 2000) and by the outputs they produce (Shneiderman, 2020; Chassang et al., 2021). Figure 3 shows each role’s picks by output and function in the taxonomy of the Method.

Tool function differed by role ($\chi^2(14, N = 1{,}568) = 77.26$, $V = 0.157$). A designer-level permutation test agreed ($p < 0.001$). By phase, the role gap held in Discover, Define and Develop (Holm-adjusted permutation $p \leq 0.002$) and was largest in Develop ($V = 0.242$). Deliver showed no significant difference ($V = 0.175$, Holm-adjusted permutation $p = 0.156$). Conversational tools took almost half of every role’s picks (47.6% to 48.1%). Industrial Designers had the largest shares of AI-native image generation (20.9%) and of CAD platforms with embedded AI (4.8%). Most of their image generation picks fell in Develop (67 of 92, 72.8%). Midjourney was their most picked image generator (31 picks). UX/UI Designers had the largest share of design software with embedded AI (29.1%), followed by Product Designers (25.2%). Figma AI led this function in both roles (58 picks each), with FigJam AI ahead in Define and Adobe Photoshop in Deliver (see tables in appendices 5 to 7). Among image, video and 3D picks, AI-native tools took a larger share for Industrial Designers (102 of 207, 49.3%) than for Product (87 of 251, 34.7%) and UX/UI Designers (65 of 217, 30.0%). This difference held at the designer level ($\chi^2(2, N = 675) = 18.35$, $V = 0.165$, permutation $p = 0.003$). AI-native tools clustered in Develop, where they took 47.7% to 72.3% of each role's picks, against 10.9% or less in the other phases (Appendices 5 to 7).

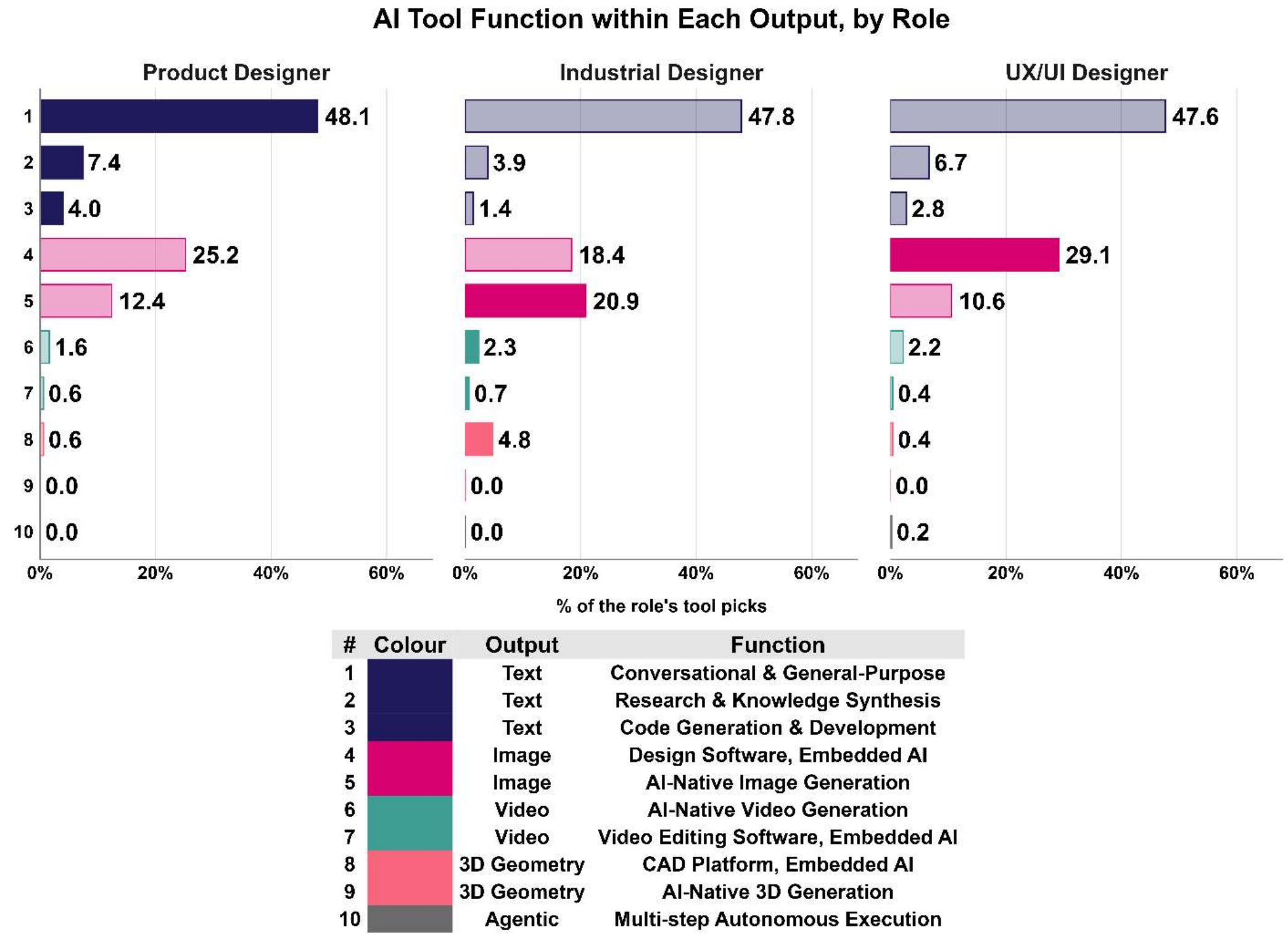

***Note. Bars give the share of each role's tool picks (620, 441 and 508). Colours follow the five outputs. Full colour marks the role with the largest share of that function.***

**Figure 3: AI Tool Function within Each Output, by Role**

## *Common, Shared and Exclusive Tools*

Toolkit size per AI adopter did not differ significantly by role ($M = 3.91$, 3.44 and 3.94 distinct tools, $H(2) = 3.34$, $p = 0.188$). Of the 60 tools, 25 were common to all three roles, nine were shared by two roles and 26 were exclusive to one (Product 10, Industrial 11, UX/UI 5). The common tools carried 579 of 620 Product picks (93.4%), 397 of 441 Industrial picks (90.0%) and 486 of 508 UX/UI picks (95.7%). Figure 4 shows this split within each phase. The mix of common and role-only tools did not differ significantly across phases ($N = 107$ tool-phase entries, shared tools excluded, and confirm 12.64 against the 4-decimal print, $p = 0.179$, Fisher's exact $p = 0.177$). UX/UI Designers had no exclusive tool in Discover and one in Define, against five in Develop (22.7%) and four in Deliver (30.8%). Industrial Designers had their largest exclusive share in Define (10 of 20 tools, 50.0%). Product Designers' exclusive share ranged from 29.4% to 40.7% and peaked in Develop (Figure 4).

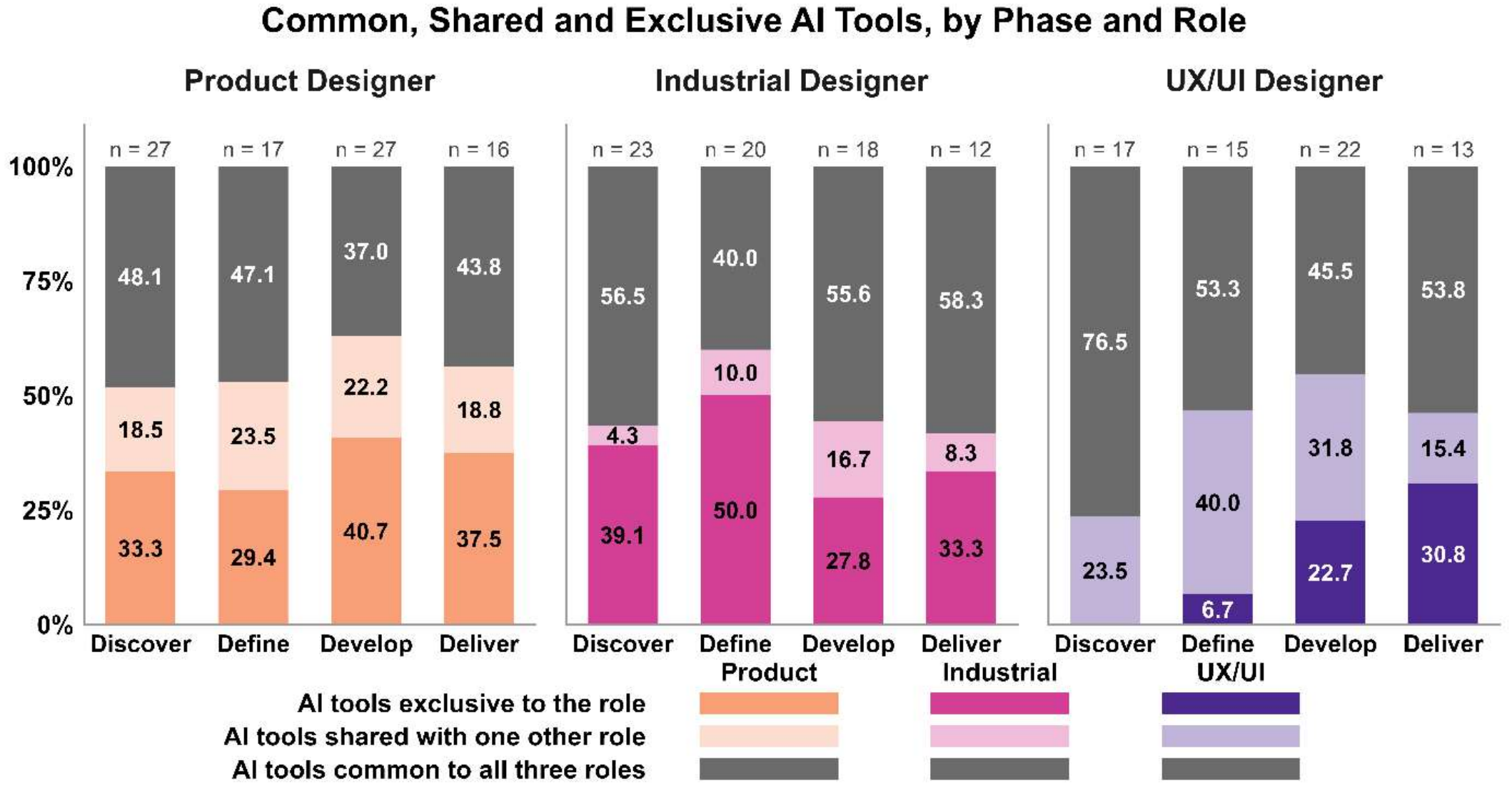

***Note. n above each bar = distinct AI tools of the role in that phase.***

**Figure 4: Common, Shared and Exclusive AI Tools, by Phase and Role**

## *Tool Loyalty*

Table 5 lists the loyal tools of each role, eight for Product Designers, three for Industrial Designers and four for UX/UI Designers. Only ChatGPT and Gemini were loyal for all three roles. Most loyal tools were **Conversational & General-Purpose**. The exceptions were NotebookLM, Figma Make and Miro AI among Product Designers and Nano Banana among Industrial Designers. Loyal tools took 47.5% to 67.1% of picks in Discover, Define and Deliver, and 13.9% to 19.5% in Develop (Figure 5). ChatGPT was not offered in Develop (Table 2), so part of this dip may follow the form. All three roles drew on one shared core of tools and differed in function shares and in the few tools they kept in every phase. The third question concerned intensity in each phase.

**Table 5: Loyal AI Tools and Their Share of Picks, by Phase and Role**

| Role | Loyal AI tool | Discover | Define | Develop | Deliver | Total picks | Offered in |
|---|---|---|---|---|---|---|---|
| **Product** | ChatGPT | 86 | 82 | 10 | 54 | 232 | Discover, Define, Deliver |
| | Gemini | 14 | 10 | 6 | 6 | 36 | none |
| | Miro AI | 15 | 14 | 2 | 1 | 32 | Discover, Define |
| | Claude | 6 | 5 | 1 | 2 | 14 | none |
| | Figma Make | 1 | 1 | 5 | 1 | 8 | none |
| | Grok | 3 | 1 | 3 | 1 | 8 | none |
| | Microsoft Copilot | 3 | 1 | 1 | 1 | 6 | none |
| | NotebookLM | 2 | 2 | 1 | 1 | 6 | none |
| **Industrial** | ChatGPT | 71 | 60 | 5 | 33 | 169 | Discover, Define, Deliver |
| | Gemini | 13 | 7 | 6 | 4 | 30 | none |
| | Nano Banana | 2 | 1 | 3 | 1 | 7 | none |
| **UX/UI** | ChatGPT | 77 | 75 | 7 | 39 | 198 | Discover, Define, Deliver |
| | Gemini | 11 | 9 | 6 | 3 | 29 | none |
| | Microsoft Copilot | 2 | 1 | 2 | 2 | 7 | none |
| | Grok | 2 | 1 | 1 | 1 | 5 | none |

***Note. Loyal = picked by the role in all four phases. Phase columns give the designers of the role who picked the tool in that phase. Offered in = phases where the tool was a tick box.***

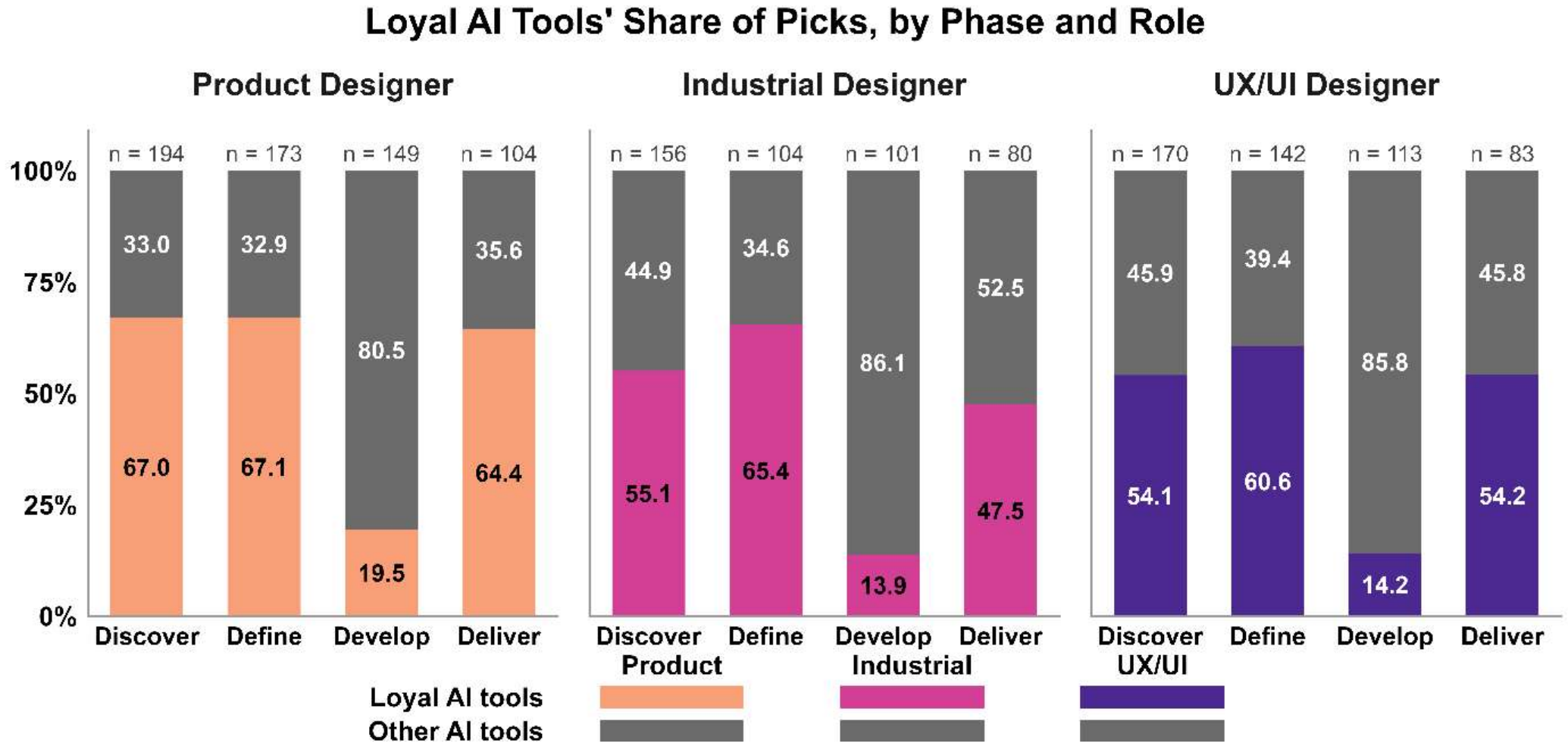

***Note. n above each bar = tool picks of the role in that phase. Loyal AI tools = tools the role picked in all four phases. ChatGPT was not offered in Develop, and its 22 Develop picks were typed in.***

**Figure 5: Loyal AI Tools' Share of Picks, by Phase and Role**

## *The Depth of Engagement across the Workflow*

Each designer gave a tool count in every phase, so Friedman tests compared the phases within designers. Intensity fell across the four phases in every role ($\chi^2(3) = 37.10$, 30.15 and 43.95, all $p < 0.001$). Kendall's W was largest for UX/UI Designers (0.15) and smallest for Industrial Designers (0.08). Among phase AI adopters, Industrial and UX/UI Designers named fewer tools in Define than in Discover and more in Develop (Figure 6). Product Designers showed no such rise (M = 1.97 in Define and 1.96 in Develop). Without the 12 Develop adopters who named no identifiable tool, the Develop means rose only slightly (Product 2.13, Industrial 1.94, UX/UI 2.51). Roles differed in intensity in Discover ($H(2) = 10.17$, Holm-adjusted $p = 0.019$) and Define ($H(2) = 19.94$, Holm-adjusted $p < 0.001$). Industrial Designers had the lowest means in both phases, at 1.31 tools in Discover and 0.87 in Define across all designers. In Develop and Deliver, the roles did not differ significantly (both Holm-adjusted $p = 0.091$). Wilcoxon signed-rank tests, which compare two phases within the same designers, traced a different path in each role (Appendix Table 8). Product Designers declined in steps, since every pair of phases differed except Discover with Define and Define with Develop (Holm-adjusted $p \leq 0.012$). Industrial Designers dropped after Discover (all Holm-adjusted $p < 0.01$) and then stayed level (Holm-adjusted $p = 0.133$ to $0.997$). UX/UI Designers named more tools in Discover than in any later phase (Holm-adjusted $p \leq 0.036$), and more in Define than in Deliver (Holm-adjusted $p < 0.001$).

These rank tests could not tell whether role differences lay in using AI in a phase or in the number of tools named. In the hurdle model, the Phase × Role interaction was significant for adoption ($\chi^2(6) = 13.12$, $p = 0.041$) and was not shown for the number of tools ($\chi^2(6) = 11.27$, $p = 0.080$). Only one interaction term was significant, UX/UI Designers in Develop (OR = 0.26, $p = 0.017$). For a typical UX/UI Designer, predicted adoption fell from 87.2% in Define to 43.9% in Develop (Figure 7). The fall over the same step was smaller for Product Designers (85.5% to 72.7%) and Industrial Designers (58.7% to 45.4%). Role differences in intensity were thus found for whether designers used AI in a phase, and not shown for how many tools they named once they did. The Discussion compares these patterns with earlier studies.

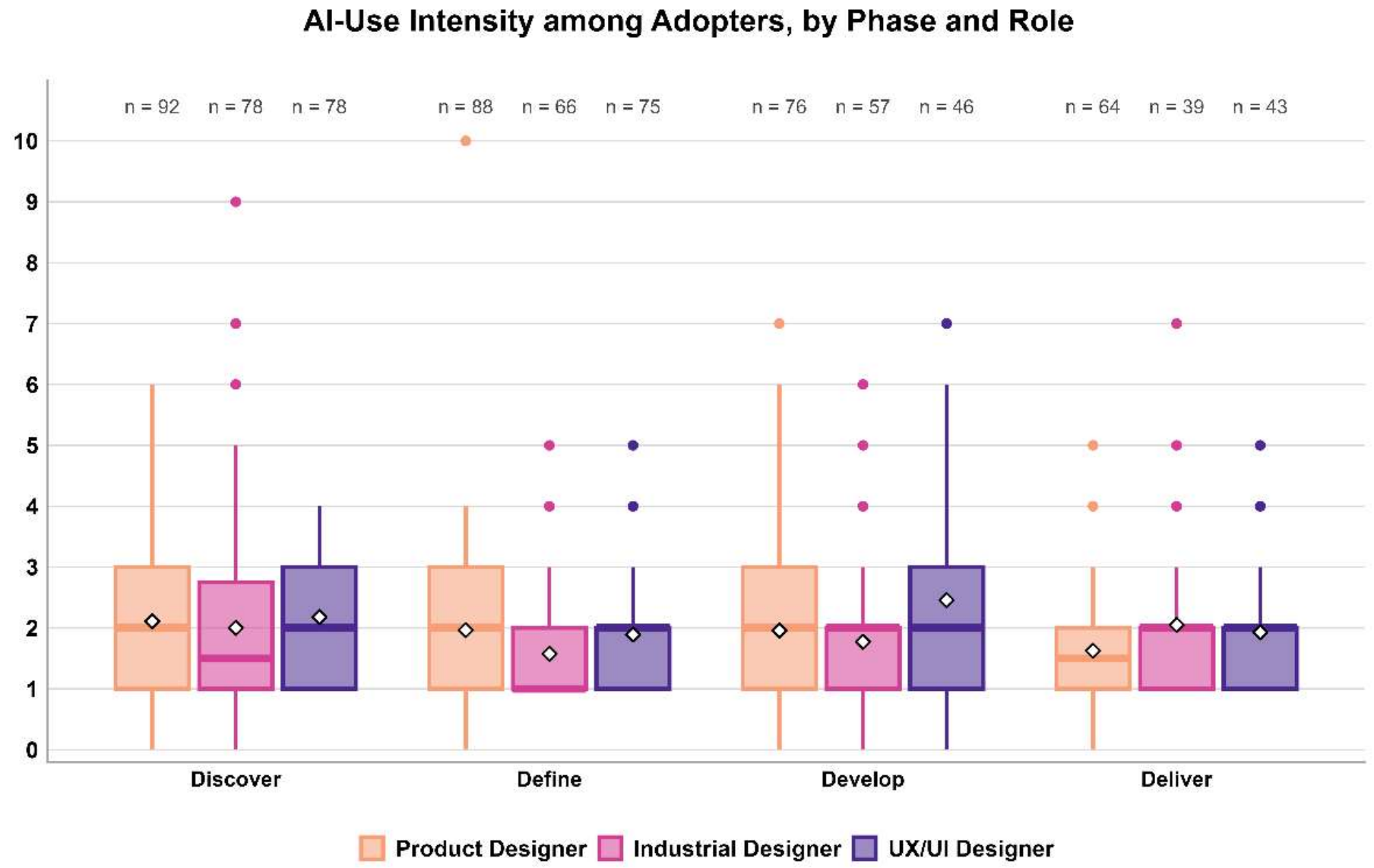

***Note. Box = middle 50%. Diamond = mean. Dots = outliers. n = phase AI adopters. A value of 0 = "Yes" in the phase with no identifiable AI tool (12 in Develop). In four boxes the median sits on the upper edge.***

**Figure 6: AI-Use Intensity among Adopters, by Phase and Role**

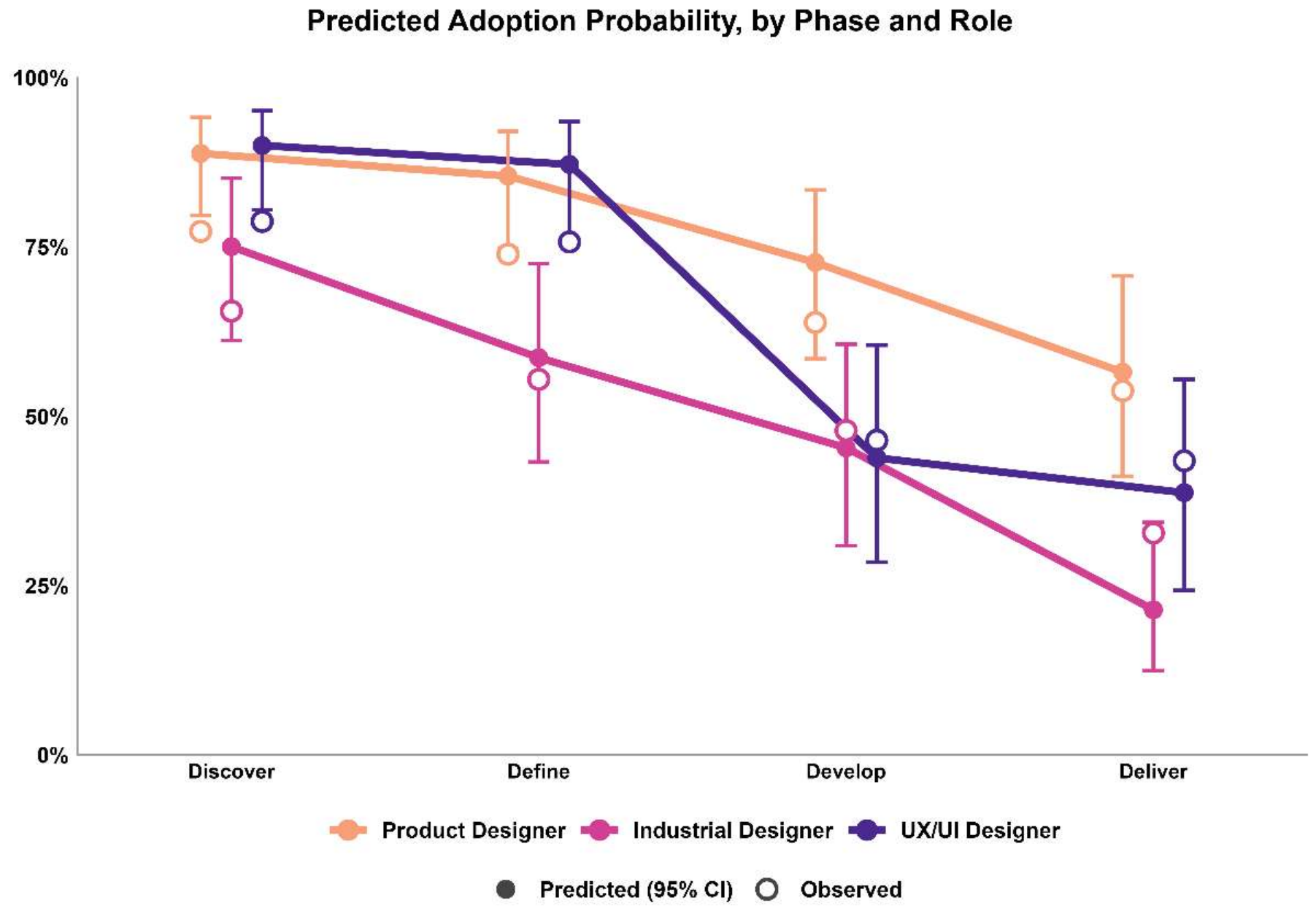

***Note. Filled points = predicted adoption for a typical designer (random effect 0) with 95% CI, from Part 1 of the hurdle model. Open circles = observed shares. n = 119, 119 and 99.***

**Figure 7: Predicted Adoption Probability, by Phase and Role**

# DISCUSSION

The three roles drew on one shared AI toolkit and differed at its edges. Tools common to all three roles carried nine of every ten picks, and toolkits were of similar size. Function use set the roles apart. In a field experiment, consultants with AI did better on tasks inside its capability frontier and worse beyond it (Dell'Acqua et al., 2026). Roles may thus share AI where their tasks overlap and part where their outputs differ. This comparison answers the call for studies by design area (Fang, Zhang, et al., 2026). The first edge concerned whether designers adopted AI at all.

Industrial Designers adopted AI less often, yet the gap went with physical projects once project type was held constant. Role and project type overlapped, so the two cannot be fully separated. Industrial designers describe a pull between automated image making and their own design judgment, and a risk of losing control over small decisions (Süner-Pla-Cerdà et al., 2026; Abrahamsen & Sjödell, 2026). AI resources are also strongest for text and images and weakest for 3D modelling (Veggi, 2025). Generative design tools ask designers to frame constraints first (Saadi & Yang, 2023), and support agents helped mechanical designers reach feasible designs (Gmeiner et al., 2025). Physical work may thus offer fewer tasks where current tools fit. How often designers used AI showed a related split.

Product and UX/UI Designers used AI daily about twice as often as Industrial Designers, and frequent users in every role spread AI over more phases. Habit grows from frequent and wide use (Limayem et al., 2007) and predicts later use (Venkatesh et al., 2012). Designers in the pooled sample also added phases in workflow order, much as innovations diffuse (Rogers, 2003; Tajarmakan & Hojjati Emami, 2026). Experience showed no significant link with adoption in the adjusted model (Appendix Table 4). Among developers too, years of coding shaped the role given to AI more than its uptake (Zakharov et al., 2025). Danish workers differed, since younger and less experienced workers adopted ChatGPT more often (Humlum & Vestergaard, 2025). One reason may be its overlap with role and project type here. The tools themselves added a further layer.

Each role leaned on tools that suit its own medium. Industrial Designers drew more on AI-native image generation and CAD, and UX/UI Designers on embedded AI. An industrial designer likened Midjourney to Pinterest for early exploration (Uusitalo et al., 2024), while UX practitioners asked for generative UI inside tools such as Figma (X. A. Chen et al., 2025). The roles differed most in Develop, where built systems cluster (Lee et al., 2025) and designers move from directing a conversation toward curating generated output (Guo et al., 2023; He et al., 2024). Develop was also where AI use among UX/UI Designers dropped most steeply. ChatGPT was weak at prototyping (Z. Zhou et al., 2024), and checking AI output can absorb the time it saves (Simkute et al., 2024). Once designers used AI in a phase, the number of tools differed little by role. Each phase thus posed its own adoption decision (Tajarmakan & Hojjati Emami, 2026). The tools kept across all phases complete the picture.

A few loyal tools carried about half of each role's picks, and ChatGPT gave most of them. Habit with a tool already in use builds inertia against switching (Polites & Karahanna, 2012), and satisfaction and usefulness keep a system in use (Bhattacherjee, 2001). Loyalty here was behavioral, since repeat use alone is no proof of commitment (Oliver, 1999), and tool sets change over time (Bødker & Klokmose, 2012). Product Designers kept the widest loyal set. One reason may be that their title spans physical and digital work (Rosala & Krause, 2019; Chochinov, 2023), so their toolkit may combine both. The loyal share fell in Develop, where ChatGPT was not a tick box, and check-all lists draw fewer choices than forced-choice items (Smyth et al., 2006). These patterns point to concrete steps for practice.

### *Practical implications*

Tool builders can bring AI into the design software each role already uses, as UX practitioners asked when they tested generative UI (X. A. Chen et al., 2025). Industrial Designers would gain from support for physical constraints and 3D work (Veggi, 2025). Design teams can plan one shared core of tools and add role-specific tools by phase. Training belongs in that plan, as employer rules and training needs held back ChatGPT use among Danish workers (Humlum & Vestergaard, 2025). Educators can teach AI literacy, the competence to judge AI critically and work with it (Long & Magerko, 2020). Junior designers deserve attention, since they may lose skills (J. Li et al., 2024). These links are associations, and each team can check them in its own practice.

### *Limitations*

Reach across real projects came at the cost of experimental control. The design was cross-sectional and self-reported, so the results are associations, and the latest project may not match a designer's typical work. Role and project type overlapped, and the title product designer covers physical and digital work (Rosala & Krause, 2019; Chochinov, 2023), which may blur role differences. The form listed between five and seven tools per phase, and ChatGPT was not listed in Develop. Check-all lists draw fewer choices (Smyth et al., 2006), so typed-in tools and the Develop loyal share may be under-counted. Loyalty was behavioral, and one designer per phase could make a tool loyal, which favors the role with the most adopters. Seventeen phase answers named no identifiable tool and counted as zero, which lowers Develop intensity slightly. Snowball sharing drew 43.0% of designers from Iran, with the largest shares among Product and Industrial Designers. Access to AI tools may be narrower there, as access moderated adoption in another national sample (Fang, Zhang, et al., 2026), which may lower adoption among Product and Industrial Designers most. One-phase groups of nine to 11 designers were only described, and graphic designers, students and other roles were left out. Tool markets move quickly, and the tool patterns of 2025 and 2026 may date soon.

### *Suggestions for Future Research*

These limits suggest four designs. A diary or log study could track one project from Discover to Deliver, as log and journal studies of UX practitioners did over weeks (Z. Zhou et al., 2024; X. A. Chen et al., 2025). Surveying the same panel a year later would show whether loyal sets persist and when designers switch tools (Polites & Karahanna, 2012; Limayem et al., 2007). Forced-choice tool items of equal length in every phase can remove the list effect. Samples balanced by project type within each role can separate role from medium, and adding graphic designers and students widens the comparison.

## CONCLUSION

Across 40 countries and regions, 337 designers in three roles reported their AI use phase by phase on a recent project. The three roles shared one core of AI tools and differed at its edges. Industrial

Designers adopted AI less often, but the gap went with physical projects once project type and country were held constant. Product and UX/UI Designers used AI daily about twice as often, and frequent users in every role spread AI over more phases. The roles also leaned on different kinds of tools, and the gap was widest in Develop. There, Industrial Designers turned to AI-native image generation, while UX/UI Designers turned to AI inside the design software they already use. A few loyal tools carried about half of each role's picks, with ChatGPT in the lead. Product Designers kept the widest loyal set. The number of tools fell across the phases in every role. Role differences lay in whether designers used AI in a phase, most clearly in the steep drop among UX/UI Designers from Define to Develop. AI use thus has a common base and role-specific parts. Tool makers can build for the medium of each role, and teams can pair one shared core with role tools. Educators can prepare students to choose and judge AI tools for their own kind of design work.

## *Acknowledgments*

The authors thank the 443 designers from at least 43 countries and regions who took part in this study. They also thank the design communities and networks that helped distribute the survey.

## *CRediT Author Statement*

Sepideh Tajarmakan: Conceptualization, Methodology, Investigation, Data curation, Formal analysis, Software, Visualization, Writing – original draft.

Khashayar Hojjati Emami: Supervision, Validation, Writing – review & editing.

## *Consent and Ethical Considerations*

The questionnaire's first item (Q1) presented an informed consent statement. It read, "By continuing, you agree to participate in this anonymous research study on AI tools in design workflows. Your responses will be confidential and used only for academic purposes." Participants selected "Yes" or "No" to "I consent to participate", and only "Yes" respondents proceeded. Participation was voluntary, and participants could withdraw at any point before submission. No names, telephone numbers or other personal details were collected. An optional final item invited an email address for a possible follow-up interview. No interviews took place. All email addresses were removed before analysis, and the deposited files contain no personal data. The original export is held by the Department of Industrial Design, Iran University of Art, and is not shared. This department reviewed the study and confirmed in writing that it did not require an ethics approval code.

## *Data Availability Statement*

The analysis code (R), README and codebook are available on Zenodo: https://doi.org/10.5281/zenodo.24239213. The cleaned master dataset, and the tool taxonomy are deposited in a second record: https://doi.org/10.5281/zenodo.24241427. The data come from the same survey as the companion paper (Tajarmakan & Hojjati Emami, 2026). Access to the data record is restricted, because the dataset supports an ongoing paper series, requests go to the corresponding author and are granted case by case. The files will be opened when the series is complete. Running the code with these files reproduces all tables and figures.

### Declaration of Generative AI Use

Grammarly was used for grammar and language editing. Claude (Anthropic) was used to refine figure design. All data analysis, statistical testing and interpretation were conducted by the authors, who reviewed and edited all text and take full responsibility for the content.

### Funding

This work received no external funding.

### Declaration of Interest Statement

The authors report there are no competing interests to declare.

# APPENDIX

## Appendix Table 1: Recognition Effect between General and Phase-Specific AI-Use Items, by Phase

| Phase | Phase named in Q12 | "Yes" in the phase question | "Yes" in the phase, not in Q12 | In Q12, not "Yes" in the phase | $\chi^2(1)$ | p |
|---|---|---|---|---|---|---|
| **Discover** | 210 | 248 | 50 | 12 | 22.08 | < 0.001 |
| **Define** | 191 | 229 | 54 | 16 | 19.56 | < 0.001 |
| **Develop** | 160 | 179 | 48 | 29 | 4.21 | 0.040 |
| **Deliver** | 86 | 146 | 74 | 14 | 39.56 | < 0.001 |

*Note. McNemar tests with continuity correction. Q12 asked in which steps of design work the designer used AI. N = 337.*

## Appendix Table 2: Example of Binary Tool Coding (n = 3 Participants)

| Participant | Open-text response | ChatGPT | Grok | Gemini | Siemens NX | Altair Inspire |
|---|---|---|---|---|---|---|
| **1** | Chatgpt | 1 | 0 | 0 | 0 | 0 |
| **2** | Grok | 0 | 1 | 0 | 0 | 0 |
| **3** | Siemens NX, Altair Inspire | 0 | 0 | 0 | 1 | 1 |

***Note.*** **A tool named or ticked was coded 1 and every other tool column 0, in the same way for ticked and typed tools and in all four phases.**

## Appendix Table 3: Country or Region by Role (N = 337)

| | Product | | Industrial | | UX/UI | | Total | |
|---|---|---|---|---|---|---|---|---|
| **Country or region** | **n** | **%** | **n** | **%** | **n** | **%** | **n** | **%** |
| Iran | 63 | 52.9 | 54 | 45.4 | 28 | 28.3 | 145 | 43.0 |
| India | 9 | 7.6 | 11 | 9.2 | 10 | 10.1 | 30 | 8.9 |
| Italy | 4 | 3.4 | 10 | 8.4 | 12 | 12.1 | 26 | 7.7 |
| Turkiye | 2 | 1.7 | 9 | 7.6 | 9 | 9.1 | 20 | 5.9 |
| United States | 9 | 7.6 | 5 | 4.2 | 3 | 3.0 | 17 | 5.0 |
| Germany | 6 | 5.0 | 2 | 1.7 | 3 | 3.0 | 11 | 3.3 |
| Canada | 1 | 0.8 | 3 | 2.5 | 6 | 6.1 | 10 | 3.0 |
| Sri Lanka | 1 | 0.8 | 0 | 0.0 | 6 | 6.1 | 7 | 2.1 |
| Spain | 2 | 1.7 | 2 | 1.7 | 2 | 2.0 | 6 | 1.8 |
| Australia | 3 | 2.5 | 0 | 0.0 | 1 | 1.0 | 4 | 1.2 |
| Egypt | 3 | 2.5 | 1 | 0.8 | 0 | 0.0 | 4 | 1.2 |
| France | 2 | 1.7 | 1 | 0.8 | 1 | 1.0 | 4 | 1.2 |
| Pakistan | 2 | 1.7 | 0 | 0.0 | 2 | 2.0 | 4 | 1.2 |
| Portugal | 2 | 1.7 | 1 | 0.8 | 1 | 1.0 | 4 | 1.2 |
| South Africa | 0 | 0.0 | 1 | 0.8 | 3 | 3.0 | 4 | 1.2 |
| United Kingdom | 1 | 0.8 | 1 | 0.8 | 2 | 2.0 | 4 | 1.2 |
| Argentina | 0 | 0.0 | 2 | 1.7 | 0 | 0.0 | 2 | 0.6 |
| Bosnia and Herzegovina | 0 | 0.0 | 1 | 0.8 | 1 | 1.0 | 2 | 0.6 |
| Indonesia | 1 | 0.8 | 1 | 0.8 | 0 | 0.0 | 2 | 0.6 |
| Nigeria | 1 | 0.8 | 0 | 0.0 | 1 | 1.0 | 2 | 0.6 |
| Sweden | 1 | 0.8 | 0 | 0.0 | 1 | 1.0 | 2 | 0.6 |
| Bangladesh | 0 | 0.0 | 0 | 0.0 | 1 | 1.0 | 1 | 0.3 |
| Belgium | 0 | 0.0 | 0 | 0.0 | 1 | 1.0 | 1 | 0.3 |
| Ethiopia | 0 | 0.0 | 1 | 0.8 | 0 | 0.0 | 1 | 0.3 |
| Finland | 1 | 0.8 | 0 | 0.0 | 0 | 0.0 | 1 | 0.3 |
| Hong Kong | 0 | 0.0 | 1 | 0.8 | 0 | 0.0 | 1 | 0.3 |
| Ireland | 0 | 0.0 | 1 | 0.8 | 0 | 0.0 | 1 | 0.3 |
| Japan | 0 | 0.0 | 0 | 0.0 | 1 | 1.0 | 1 | 0.3 |
| Jordan | 1 | 0.8 | 0 | 0.0 | 0 | 0.0 | 1 | 0.3 |
| Kenya | 0 | 0.0 | 1 | 0.8 | 0 | 0.0 | 1 | 0.3 |
| Malaysia | 0 | 0.0 | 1 | 0.8 | 0 | 0.0 | 1 | 0.3 |
| Mexico | 0 | 0.0 | 1 | 0.8 | 0 | 0.0 | 1 | 0.3 |
| Netherlands | 1 | 0.8 | 0 | 0.0 | 0 | 0.0 | 1 | 0.3 |
| New Zealand | 1 | 0.8 | 0 | 0.0 | 0 | 0.0 | 1 | 0.3 |
| Peru | 0 | 0.0 | 1 | 0.8 | 0 | 0.0 | 1 | 0.3 |
| Russia | 0 | 0.0 | 1 | 0.8 | 0 | 0.0 | 1 | 0.3 |
| South Korea | 0 | 0.0 | 1 | 0.8 | 0 | 0.0 | 1 | 0.3 |
| Switzerland | 0 | 0.0 | 1 | 0.8 | 0 | 0.0 | 1 | 0.3 |
| Taiwan | 0 | 0.0 | 0 | 0.0 | 1 | 1.0 | 1 | 0.3 |
| Vietnam | 0 | 0.0 | 0 | 0.0 | 1 | 1.0 | 1 | 0.3 |
| **Prefer not to say** | 2 | 1.7 | 4 | 3.4 | 2 | 2.0 | 8 | 2.4 |

***Note. Hong Kong and Taiwan are regions. "Prefer not to say" = country not answered (n = 8). Product Designers came from 22 countries and regions, Industrial Designers from 26 and UX/UI Designers from 23.***

## Appendix Table 4: Adjusted Model of AI Adoption

***Panel A. Odds ratios (logistic regression, n = 334 designers, 278 AI adopters)***

| **Term** | **b** | **OR** | **95% CI** | **p** |
|---|---|---|---|---|
| Industrial Designer (vs Product Designer) | −0.643 | 0.53 | [0.24, 1.17] | 0.115 |
| UX/UI Designer (vs Product Designer) | −0.673 | 0.51 | [0.20, 1.30] | 0.158 |
| Hybrid project (vs Digital) | −0.557 | 0.57 | [0.24, 1.35] | 0.203 |
| Physical project (vs Digital) | −1.383 | 0.25 | [0.10, 0.63] | 0.003 |
| Experience (one level up) | −0.282 | 0.75 | [0.53, 1.08] | 0.122 |
| Education (one level up) | 0.411 | 1.51 | [0.90, 2.52] | 0.116 |
| Iran (vs other countries) | −0.809 | 0.45 | [0.24, 0.83] | 0.010 |

***Panel B. Likelihood ratio test per variable***

| **Variable** | **df** | **$\chi^2$** | **p** |
|---|---|---|---|
| Role | 2 | 3.37 | 0.185 |
| Project type | 2 | 9.38 | 0.009 |
| Experience | 1 | 2.43 | 0.119 |
| Education | 1 | 2.53 | 0.112 |
| Iran | 1 | 6.67 | 0.010 |

***Note. Education "Other" (n = 3) was left out. Experience and education entered as ranks (one level up). Iran = designers in Iran against all other countries. 95% CI = Wald interval.***

## Appendix Table 5: AI Tool Names by Output, Function and Phase, Product Designers

| | | | **Discover** | | **Define** | | **Develop** | | **Deliver** | | |
|---|---|---|---|---|---|---|---|---|---|---|---|
| **Output** | **Function** | **AI tool** | **n** | **%** | **n** | **%** | **n** | **%** | **n** | **%** | **Total** |
| **Text** | Conversational & General-Purpose | **ChatGPT*** | 86 | 44.3 | 82 | 47.4 | 10 | 6.7 | 54 | 51.9 | 232 |
| | | **Gemini** | 14 | 7.2 | 10 | 5.8 | 6 | 4.0 | 6 | 5.8 | 36 |
| | | **Claude** | 6 | 3.1 | 5 | 2.9 | 1 | 0.7 | 2 | 1.9 | 14 |
| | | **Grok** | 3 | 1.5 | 1 | 0.6 | 3 | 2.0 | 1 | 1.0 | 8 |
| | | **Microsoft Copilot** | 3 | 1.5 | 1 | 0.6 | 1 | 0.7 | 1 | 1.0 | 6 |
| | | **Kimi** | 1 | 0.5 | 1 | 0.6 | 0 | 0.0 | 0 | 0.0 | 2 |
| | Research & Knowledge Synthesis | **Perplexity AI*** | 17 | 8.8 | 0 | 0.0 | 1 | 0.7 | 0 | 0.0 | 18 |
| | | **NotebookLM** | 2 | 1.0 | 2 | 1.2 | 1 | 0.7 | 1 | 1.0 | 6 |
| | | **Dovetail** | 2 | 1.0 | 1 | 0.6 | 0 | 0.0 | 0 | 0.0 | 3 |
| | | **Notion AI*** | 1 | 0.5 | 15 | 8.7 | 0 | 0.0 | 1 | 1.0 | 17 |
| | | **Consensus** | 1 | 0.5 | 0 | 0.0 | 0 | 0.0 | 0 | 0.0 | 1 |
| | | **Elicit*** | 1 | 0.5 | 0 | 0.0 | 0 | 0.0 | 0 | 0.0 | 1 |
| | Code Generation & Development | **Lovable** | 2 | 1.0 | 1 | 0.6 | 2 | 1.3 | 0 | 0.0 | 5 |
| | | **Figma Make** | 1 | 0.5 | 1 | 0.6 | 5 | 3.4 | 1 | 1.0 | 8 |
| | | **Bolt.new** | 1 | 0.5 | 0 | 0.0 | 2 | 1.3 | 1 | 1.0 | 4 |
| | | **Cursor** | 1 | 0.5 | 0 | 0.0 | 1 | 0.7 | 0 | 0.0 | 2 |
| | | **Framer AI** | 1 | 0.5 | 0 | 0.0 | 0 | 0.0 | 0 | 0.0 | 1 |
| | | **UX Pilot** | 1 | 0.5 | 0 | 0.0 | 0 | 0.0 | 0 | 0.0 | 1 |
| | | **Google Stitch** | 0 | 0.0 | 0 | 0.0 | 2 | 1.3 | 0 | 0.0 | 2 |
| | | **Replit** | 0 | 0.0 | 0 | 0.0 | 1 | 0.7 | 0 | 0.0 | 1 |
| | | **V0** | 0 | 0.0 | 0 | 0.0 | 1 | 0.7 | 0 | 0.0 | 1 |
| **Image** | Design Software, Embedded AI | **Figma AI*** | 20 | 10.3 | 0 | 0.0 | 36 | 24.2 | 2 | 1.9 | 58 |
| | | **Miro AI*** | 15 | 7.7 | 14 | 8.1 | 2 | 1.3 | 1 | 1.0 | 32 |
| | | **Adobe Sensei*** | 3 | 1.5 | 0 | 0.0 | 0 | 0.0 | 0 | 0.0 | 3 |
| | | **Freepik** | 1 | 0.5 | 1 | 0.6 | 0 | 0.0 | 0 | 0.0 | 2 |
| | | **FigJam AI*** | 0 | 0.0 | 31 | 17.9 | 0 | 0.0 | 0 | 0.0 | 31 |
| | | **Whimsical AI*** | 0 | 0.0 | 5 | 2.9 | 1 | 0.7 | 1 | 1.0 | 7 |
| | | **Adobe Photoshop*** | 0 | 0.0 | 0 | 0.0 | 0 | 0.0 | 15 | 14.4 | 15 |
| | | **Canva AI*** | 0 | 0.0 | 0 | 0.0 | 0 | 0.0 | 8 | 7.7 | 8 |
| | AI-Native Image Generation | **Midjourney*** | 5 | 2.6 | 0 | 0.0 | 30 | 20.1 | 0 | 0.0 | 35 |
| | | **Vizcom** | 3 | 1.5 | 0 | 0.0 | 0 | 0.0 | 0 | 0.0 | 3 |
| | | **Visual Electric** | 1 | 0.5 | 1 | 0.6 | 1 | 0.7 | 0 | 0.0 | 3 |
| | | **Nano Banana** | 1 | 0.5 | 0 | 0.0 | 3 | 2.0 | 0 | 0.0 | 4 |
| | | **DALL·E*** | 0 | 0.0 | 0 | 0.0 | 21 | 14.1 | 0 | 0.0 | 21 |
| | | **Adobe Firefly*** | 0 | 0.0 | 0 | 0.0 | 9 | 6.0 | 0 | 0.0 | 9 |
| | | **Leonardo AI** | 0 | 0.0 | 0 | 0.0 | 1 | 0.7 | 0 | 0.0 | 1 |
| | | **Lovart** | 0 | 0.0 | 0 | 0.0 | 1 | 0.7 | 0 | 0.0 | 1 |
| **Video** | AI-Native Video Generation | **Runway ML*** | 0 | 0.0 | 0 | 0.0 | 4 | 2.7 | 5 | 4.8 | 9 |
| | | **Higgsfield** | 0 | 0.0 | 0 | 0.0 | 1 | 0.7 | 0 | 0.0 | 1 |
| | Video Editing Software, Embedded AI | **Descript*** | 0 | 0.0 | 0 | 0.0 | 0 | 0.0 | 4 | 3.8 | 4 |
| **3D Geometry** | CAD Platform, Embedded AI | **Autodesk Generative Design*** | 1 | 0.5 | 0 | 0.0 | 0 | 0.0 | 0 | 0.0 | 1 |
| | | **Autodesk Fusion 360 AI*** | 0 | 0.0 | 1 | 0.6 | 2 | 1.3 | 0 | 0.0 | 3 |

***Note. * = offered as a tick box in at least one phase. % = share of the role's picks in that phase (194, 173, 149 and 104).***

## Appendix Table 6: AI Tool Names by Output, Function and Phase, Industrial Designers

| | | | Discover | | Define | | Develop | | Deliver | | |
|---|---|---|---|---|---|---|---|---|---|---|---|
| **Output** | **Function** | **AI tool** | **n** | **%** | **n** | **%** | **n** | **%** | **n** | **%** | **Total** |
| **Text** | Conversational & General-Purpose | **ChatGPT*** | 71 | 45.5 | 60 | 57.7 | 5 | 5.0 | 33 | 41.2 | 169 |
| | | **Gemini** | 13 | 8.3 | 7 | 6.7 | 6 | 5.9 | 4 | 5.0 | 30 |
| | | **DeepSeek** | 3 | 1.9 | 2 | 1.9 | 0 | 0.0 | 0 | 0.0 | 5 |
| | | **Microsoft Copilot** | 2 | 1.3 | 0 | 0.0 | 0 | 0.0 | 0 | 0.0 | 2 |
| | | **Grok** | 1 | 0.6 | 0 | 0.0 | 1 | 1.0 | 1 | 1.2 | 3 |
| | | **Claude** | 1 | 0.6 | 0 | 0.0 | 0 | 0.0 | 0 | 0.0 | 1 |
| | | **Qwen** | 0 | 0.0 | 1 | 1.0 | 0 | 0.0 | 0 | 0.0 | 1 |
| | Research & Knowledge Synthesis | **Perplexity AI*** | 9 | 5.8 | 0 | 0.0 | 0 | 0.0 | 0 | 0.0 | 9 |
| | | **Elicit*** | 3 | 1.9 | 0 | 0.0 | 0 | 0.0 | 0 | 0.0 | 3 |
| | | **Notion AI*** | 0 | 0.0 | 3 | 2.9 | 0 | 0.0 | 0 | 0.0 | 3 |
| | | **NotebookLM** | 0 | 0.0 | 1 | 1.0 | 0 | 0.0 | 1 | 1.2 | 2 |
| | Code Generation & Development | **Google AI Studio** | 2 | 1.3 | 1 | 1.0 | 2 | 2.0 | 0 | 0.0 | 5 |
| | | **GitHub Copilot** | 0 | 0.0 | 0 | 0.0 | 0 | 0.0 | 1 | 1.2 | 1 |
| **Image** | Design Software, Embedded AI | **Figma AI*** | 9 | 5.8 | 0 | 0.0 | 6 | 5.9 | 0 | 0.0 | 15 |
| | | **Adobe Sensei*** | 8 | 5.1 | 0 | 0.0 | 0 | 0.0 | 0 | 0.0 | 8 |
| | | **Miro AI*** | 6 | 3.8 | 8 | 7.7 | 0 | 0.0 | 0 | 0.0 | 14 |
| | | **Canva AI*** | 1 | 0.6 | 1 | 1.0 | 0 | 0.0 | 11 | 13.8 | 13 |
| | | **FigJam AI*** | 0 | 0.0 | 5 | 4.8 | 0 | 0.0 | 0 | 0.0 | 5 |
| | | **Whimsical AI*** | 0 | 0.0 | 2 | 1.9 | 0 | 0.0 | 0 | 0.0 | 2 |
| | | **Adobe Photoshop*** | 0 | 0.0 | 1 | 1.0 | 1 | 1.0 | 21 | 26.2 | 23 |
| | | **Mural AI** | 0 | 0.0 | 1 | 1.0 | 0 | 0.0 | 0 | 0.0 | 1 |
| | AI-Native Image Generation | **Vizcom** | 7 | 4.5 | 3 | 2.9 | 8 | 7.9 | 0 | 0.0 | 18 |
| | | **Krea** | 3 | 1.9 | 1 | 1.0 | 0 | 0.0 | 0 | 0.0 | 4 |
| | | **Midjourney*** | 2 | 1.3 | 1 | 1.0 | 28 | 27.7 | 0 | 0.0 | 31 |
| | | **Nano Banana** | 2 | 1.3 | 1 | 1.0 | 3 | 3.0 | 1 | 1.2 | 7 |
| | | **Leonardo AI** | 1 | 0.6 | 1 | 1.0 | 1 | 1.0 | 0 | 0.0 | 3 |
| | | **Adobe Firefly*** | 0 | 0.0 | 1 | 1.0 | 12 | 11.9 | 1 | 1.2 | 14 |
| | | **DALL·E*** | 0 | 0.0 | 0 | 0.0 | 13 | 12.9 | 0 | 0.0 | 13 |
| | | **Microsoft designer** | 0 | 0.0 | 0 | 0.0 | 1 | 1.0 | 0 | 0.0 | 1 |
| | | **Whisk** | 0 | 0.0 | 0 | 0.0 | 1 | 1.0 | 0 | 0.0 | 1 |
| **Video** | AI-Native Video Generation | **Hailuo AI** | 1 | 0.6 | 0 | 0.0 | 0 | 0.0 | 0 | 0.0 | 1 |
| | | **Veo 3** | 1 | 0.6 | 0 | 0.0 | 0 | 0.0 | 0 | 0.0 | 1 |
| | | **Runway ML*** | 0 | 0.0 | 0 | 0.0 | 6 | 5.9 | 2 | 2.5 | 8 |
| | Video Editing Software, Embedded AI | **Descript*** | 0 | 0.0 | 0 | 0.0 | 0 | 0.0 | 3 | 3.8 | 3 |
| **3D Geometry** | CAD Platform, Embedded AI | **Autodesk Generative Design*** | 8 | 5.1 | 0 | 0.0 | 0 | 0.0 | 0 | 0.0 | 8 |
| | | **Siemens NX** | 1 | 0.6 | 0 | 0.0 | 1 | 1.0 | 0 | 0.0 | 2 |
| | | **Onshape** | 1 | 0.6 | 0 | 0.0 | 0 | 0.0 | 0 | 0.0 | 1 |
| | | **Autodesk Fusion 360 AI*** | 0 | 0.0 | 3 | 2.9 | 5 | 5.0 | 0 | 0.0 | 8 |
| | | **Altair Inspire** | 0 | 0.0 | 0 | 0.0 | 1 | 1.0 | 1 | 1.2 | 2 |

***Note. * = offered as a tick box in at least one phase. % = share of the role's picks in that phase (156, 104, 101 and 80).***

## Appendix Table 7: AI Tool Names by Output, Function and Phase, UX/UI Designers

| | | | Discover | | Define | | Develop | | Deliver | | |
|---|---|---|---|---|---|---|---|---|---|---|---|
| **Output** | **Function** | **AI tool** | **n** | **%** | **n** | **%** | **n** | **%** | **n** | **%** | **Total** |
| **Text** | Conversational & General-Purpose | **ChatGPT*** | 77 | 45.3 | 75 | 52.8 | 7 | 6.2 | 39 | 47.0 | 198 |
| | | **Gemini** | 11 | 6.5 | 9 | 6.3 | 6 | 5.3 | 3 | 3.6 | 29 |
| | | **Microsoft Copilot** | 2 | 1.2 | 1 | 0.7 | 2 | 1.8 | 2 | 2.4 | 7 |
| | | **Grok** | 2 | 1.2 | 1 | 0.7 | 1 | 0.9 | 1 | 1.2 | 5 |
| | | **Claude** | 2 | 1.2 | 1 | 0.7 | 0 | 0.0 | 0 | 0.0 | 3 |
| | Research & Knowledge Synthesis | **Perplexity AI*** | 20 | 11.8 | 1 | 0.7 | 0 | 0.0 | 0 | 0.0 | 21 |
| | | **Elicit*** | 2 | 1.2 | 0 | 0.0 | 0 | 0.0 | 0 | 0.0 | 2 |
| | | **NotebookLM** | 1 | 0.6 | 1 | 0.7 | 1 | 0.9 | 0 | 0.0 | 3 |
| | | **Notion AI*** | 0 | 0.0 | 8 | 5.6 | 0 | 0.0 | 0 | 0.0 | 8 |
| | Code Generation & Development | **Figma Make** | 2 | 1.2 | 0 | 0.0 | 1 | 0.9 | 1 | 1.2 | 4 |
| | | **Lovable** | 1 | 0.6 | 0 | 0.0 | 2 | 1.8 | 1 | 1.2 | 4 |
| | | **UX Pilot** | 1 | 0.6 | 0 | 0.0 | 0 | 0.0 | 0 | 0.0 | 1 |
| | | **Google AI Studio** | 0 | 0.0 | 0 | 0.0 | 1 | 0.9 | 1 | 1.2 | 2 |
| | | **Codex** | 0 | 0.0 | 0 | 0.0 | 1 | 0.9 | 0 | 0.0 | 1 |
| | | **Replit** | 0 | 0.0 | 0 | 0.0 | 1 | 0.9 | 0 | 0.0 | 1 |
| | | **Framer AI** | 0 | 0.0 | 0 | 0.0 | 0 | 0.0 | 1 | 1.2 | 1 |
| **Image** | Design Software, Embedded AI | **Figma AI*** | 27 | 15.9 | 0 | 0.0 | 31 | 27.4 | 0 | 0.0 | 58 |
| | | **Miro AI*** | 11 | 6.5 | 9 | 6.3 | 0 | 0.0 | 0 | 0.0 | 20 |
| | | **Adobe Sensei*** | 7 | 4.1 | 0 | 0.0 | 0 | 0.0 | 0 | 0.0 | 7 |
| | | **FigJam AI*** | 0 | 0.0 | 27 | 19.0 | 0 | 0.0 | 0 | 0.0 | 27 |
| | | **Whimsical AI*** | 0 | 0.0 | 5 | 3.5 | 0 | 0.0 | 0 | 0.0 | 5 |
| | | **Freepik** | 0 | 0.0 | 1 | 0.7 | 2 | 1.8 | 0 | 0.0 | 3 |
| | | **Adobe Photoshop*** | 0 | 0.0 | 0 | 0.0 | 0 | 0.0 | 19 | 22.9 | 19 |
| | | **Canva AI*** | 0 | 0.0 | 0 | 0.0 | 0 | 0.0 | 9 | 10.8 | 9 |
| | AI-Native Image Generation | **Midjourney*** | 2 | 1.2 | 1 | 0.7 | 18 | 15.9 | 0 | 0.0 | 21 |
| | | **Nano Banana** | 1 | 0.6 | 1 | 0.7 | 2 | 1.8 | 0 | 0.0 | 4 |
| | | **Adobe Firefly*** | 0 | 0.0 | 0 | 0.0 | 13 | 11.5 | 0 | 0.0 | 13 |
| | | **DALL·E*** | 0 | 0.0 | 0 | 0.0 | 12 | 10.6 | 0 | 0.0 | 12 |
| | | **FLUX.1** | 0 | 0.0 | 0 | 0.0 | 1 | 0.9 | 0 | 0.0 | 1 |
| | | **Leonardo AI** | 0 | 0.0 | 0 | 0.0 | 1 | 0.9 | 0 | 0.0 | 1 |
| | | **Lummi** | 0 | 0.0 | 0 | 0.0 | 1 | 0.9 | 0 | 0.0 | 1 |
| | | **Microsoft designer** | 0 | 0.0 | 0 | 0.0 | 1 | 0.9 | 0 | 0.0 | 1 |
| **Video** | AI-Native Video Generation | **Runway ML*** | 0 | 0.0 | 0 | 0.0 | 7 | 6.2 | 3 | 3.6 | 10 |
| | | **Kling AI** | 0 | 0.0 | 0 | 0.0 | 1 | 0.9 | 0 | 0.0 | 1 |
| | Video Editing Software, Embedded AI | **Descript*** | 0 | 0.0 | 0 | 0.0 | 0 | 0.0 | 2 | 2.4 | 2 |
| **3D Geometry** | CAD Platform, Embedded AI | **Autodesk Generative Design*** | 1 | 0.6 | 0 | 0.0 | 0 | 0.0 | 0 | 0.0 | 1 |
| | | **Autodesk Fusion 360 AI*** | 0 | 0.0 | 1 | 0.7 | 0 | 0.0 | 0 | 0.0 | 1 |
| **Agentic** | Multi-step Autonomous Execution | **Marvin** | 0 | 0.0 | 0 | 0.0 | 0 | 0.0 | 1 | 1.2 | 1 |

***Note. * = offered as a tick box in at least one phase. % = share of the role's picks in that phase (170, 142, 113 and 83).***

## Appendix Table 8: Pairwise Phase Comparisons of AI-Use Intensity, by Role

| Role | Pair | M first | M second | V | p | Holm-adjusted p |
|---|---|---|---|---|---|---|
| **Product** | Discover vs Define | 1.63 | 1.45 | 1139.5 | 0.087 | 0.173 |
| | Discover vs Develop | 1.63 | 1.25 | 2277.5 | 0.003 | 0.012 |
| | Discover vs Deliver | 1.63 | 0.87 | 2316.0 | < 0.001 | < 0.001 |
| | Define vs Develop | 1.45 | 1.25 | 1601.0 | 0.098 | 0.173 |
| | Define vs Deliver | 1.45 | 0.87 | 1812.0 | < 0.001 | < 0.001 |
| | Develop vs Deliver | 1.25 | 0.87 | 1296.0 | 0.004 | 0.012 |
| **Industrial** | Discover vs Define | 1.31 | 0.87 | 864.5 | < 0.001 | 0.001 |
| | Discover vs Develop | 1.31 | 0.85 | 1439.5 | < 0.001 | 0.001 |
| | Discover vs Deliver | 1.31 | 0.67 | 2144.5 | < 0.001 | < 0.001 |
| | Define vs Develop | 0.87 | 0.85 | 975.5 | 0.997 | 0.997 |
| | Define vs Deliver | 0.87 | 0.67 | 1138.0 | 0.044 | 0.133 |
| | Develop vs Deliver | 0.85 | 0.67 | 945.0 | 0.136 | 0.273 |
| **UX/UI** | Discover vs Define | 1.72 | 1.43 | 1091.0 | 0.012 | 0.036 |
| | Discover vs Develop | 1.72 | 1.14 | 1784.5 | 0.001 | 0.005 |
| | Discover vs Deliver | 1.72 | 0.84 | 1936.5 | < 0.001 | < 0.001 |
| | Define vs Develop | 1.43 | 1.14 | 1420.5 | 0.073 | 0.085 |
| | Define vs Deliver | 1.43 | 0.84 | 1383.0 | < 0.001 | < 0.001 |
| | Develop vs Deliver | 1.14 | 0.84 | 814.0 | 0.042 | 0.085 |

***Note. Wilcoxon signed-rank tests with continuity correction, all designers of the role (n = 119, 119 and 99). Holm correction over the six pairs within each role.***